\documentclass[12pt]{article}
\usepackage{amsmath}
\usepackage{graphicx,psfrag,epsf}
\usepackage{enumerate}
\usepackage{natbib}
\usepackage{url} % not crucial - just used below for the URL 

\usepackage{listings}
\usepackage{minted}
\usepackage[T1]{fontenc}

\usepackage{tikz}
\usetikzlibrary{shapes,arrows}

\usepackage{pythonhighlight}

\newcommand{\blind}{0}

\usepackage{amssymb}

\def\D{\mathcal{D}}
\def\M{\mathcal{M}}
\def\Q{\mathbb{Q}}
\def\R{\mathbb{R}}
\def\Meterstick{Meterstick}

\begin{document}

\def\spacingset#1{\renewcommand{\baselinestretch}%
{#1}\small\normalsize} \spacingset{1}

%%%%%%%%%%%%%%%%%%%%%%%%%%%%%%%%%%%%%%%%%%%%%%%%%%%%%%%%%%%%%%%%%%%%%%%%%%%%%%

\if0\blind
{
  \title{\bf A Grammar of Data Analysis}
  \author{
    Xunmo Yang,
    Taylor Pospisil, 
    Omkar Muralidharan,
    Google, Inc. \\
    and \\
    Dennis L. Sun \\
    Google, Inc. and Stanford University}
  \maketitle
} \fi

\if1\blind
{
  \bigskip
  \bigskip
  \bigskip
  \begin{center}
    {\LARGE\bf A Grammar of Data Analysis}
\end{center}
  \medskip
} \fi

\bigskip
\begin{abstract}
This paper outlines a grammar of data analysis, as distinct from 
grammars of data manipulation, in which the primitives
are metrics and dimensions. We describe a Python implementation of this grammar 
called Meterstick, which is agnostic to the underlying data source, 
which may be a DataFrame or a SQL database.
\end{abstract}

\noindent%
{\it Keywords:} data science, grammar of data analysis
\vfill

\newpage
\spacingset{1.45} % DON'T change the spacing!
\section{Introduction}
\label{sec:intro}

This paper proposes a grammar of data analysis, in the 
vein of grammars that have been proposed for 
graphics \citep{wilkinson2006grammar,wickham2010layered}. 
Just as a grammar of language specifies the primitives 
that make up a sentence (e.g., subject, verb, predicate), 
a grammar of data analysis 
specifies the primitives that make up an analysis 
(e.g., metrics, dimensions). By identifying 
these primitives, as well as the rules for 
composing them, we can design software libraries where
those primitives are promoted to first-class citizens. 
The goal is to streamline the data analysis workflow, much as 
{\tt ggplot2} has done for data visualization. \citep{wickham2009ggplot2}

We are not the first to propose such an idea. 
Many people regard {\tt dplyr} \citep{wickham2014dplyr} as 
a kind of grammar for data analysis. While {\tt dplyr} is indeed a 
grammar, its primitives are too low a level to facilitate the 
more complex analyses we have in mind, as we argue in 
Section~\ref{subsec:comparison}. In fairness to {\tt dplyr}, it only 
claims to be a ``grammar of {\em data manipulation},'' which is 
a precursor to data analysis. In order to facilitate 
complex analyses, a good grammar of data analysis ought to 
provide a layer of abstraction above the details of 
data manipulation. That said, implementations of a grammar of 
data analysis can and should be built on top of a good grammar of 
data manipulation, such as {\tt dplyr} in R or 
{\tt pandas} in Python.

\section{A Grammar of Data Analysis}
\label{sec:grammar}

Many data analysis problems can be framed as calculating {\bf metrics} 
over {\bf dimensions}. A metric defines the computation, while a 
a dimension defines the granularity at which that computation 
should be done. 
For example:
\begin{enumerate}
\item A baseball team wants to track the distribution of pitch types 
thrown by each pitcher to each batter. The ``dimensions'' are the pitcher and the 
    batter, and the ``metric'' is a vector of probabilities 
    representing the distribution over the different pitch types.
\item After running randomized experiments to evaluate 5 proposed new website designs, an online 
retailer wants to know the change in churn rate in each region from each of the new designs. 
The ``dimension'' is the region, 
    and the ``metric'' is a vector of changes in churn rate between each of 
    the new designs and the control design.
\item \citet{card1994american} estimated the effect of 
minimum wage on employment using a natural experiment 
where the minimum wage increased in New Jersey but stayed 
the same in Pennsylvania. They measured
employment at different fast-food restaurants before 
and after the minimum wage increase. Here, the ``dimension'' 
is the restaurant, and the ``metric'' is the difference-in-difference (DID) in the employment rate.
\end{enumerate}

At a high level, a grammar of data analysis consists of just these two components, metrics and 
dimensions, which can be independently composed. 
That is, we can easily imagine calculating the 
same metric over a different set of dimensions (e.g., 
pitcher only, instead of both pitcher and batter), 
or calculating a different metric (e.g., batting average) over the same set of dimensions.
We illustrate this grammar using our implementation of it, 
an open-source Python library called \Meterstick{}:
\begin{enumerate}
    \item In the baseball example, the metric is 
    simply a count over pitch types, and the \Meterstick{} code reflects this:
\begin{python}
pitch_types = Count("pitchtype") | Distribution(over="pitchtype")
pitch_types.compute_on(df, split_by=["pitcher", "batter"])
\end{python}
    \item In the online retail example, we transform the basic churn metric by 
    the \pyth{PercentChange} operation to obtain the final metric:
    \begin{python}
churn = (Sum("lost") / Count("lost")).set_names(["churn"])
churn_change = churn | PercentChange("experiment", "control")
churn_change.compute_on(df, split_by=["region"]))
     \end{python}
    \item In the minimum wage example, we can implement the DID estimator for the average employment rate as
    \begin{python}
employment_rate = Mean("EMP")  # equals Sum("EMP") / Count("EMP") 
did = (employment_rate
      | AbsoluteChange("STATE_NAME", "PA") 
      | AbsoluteChange("PERIOD", "Before"))
did.compute_on(df)
     \end{python}
     
% Alternatively, a regression approach can be used:

%  \begin{python}
% did_reg = LinearRegression(
%     y=Mean("EMP"),
%     x=[Mean("is_NJ"), Mean("is_after"), Mean("is_NJ_and_after")],
%     group_by='row_id')
%      \end{python}
% where \pyth{is_NJ}, \pyth{is_after} and \pyth{is_NJ_and_after} are indicator variables for observations in New Jersey, the period after the minimum wage increase, and their interaction, respectively. With \pyth{row_id} as the aggregation level, the input data will be aggreg and the \pyth{Mean}s are no-ops. The coefficient of the interaction term \pyth{is_NJ_and_after} yields the DID estimate. This dual representation offers flexibility for extensions. For instance, to calculate the DID as a percentage point difference, we can replace the initial \pyth{AbsoluteChange} with \pyth{PercentChange}. Conversely, the regression framework is advantageous for incorporating additional covariates.

    Although we calculated the absolute 
    change in employment to reproduce 
    the results of \citet{card1994american}, 
    we could just have well calculated
    the percent change in employment, instead
    reporting a percentage point difference. This is not straightforward to 
    do in the regression framework 
    used by \citet{card1994american}, but it is straightforward
    in this grammar of data analysis; we would simply need to
    replace the first \pyth{AbsoluteChange} by 
    a \pyth{PercentChange}.
    
%\Meterstick{} offers alternative ways to express DID, and readily extends to more sophisticated variations, including those that account for covariates and multiple time periods. We will elaborate on these extensions in subsequent sections.

    % TODO: An example showing standard errors for a percentage point change. Xunmo has changed the example for Bootstrap below to DID.
    
\end{enumerate}
In each case, a metric is defined and piped to the function 
\pyth{compute_on()} that finally computes that metric on data. The dimensions are specified in 
the \pyth{split_by=} argument of \pyth{compute_on()}. If we decide to change the dimensions 
later, the code only needs to be modified in one place, making the code more modular.

Compared to dimensions, metrics are much richer in their possibilities. The 
remainder of this section demonstrates how the algebraic structure of metrics 
can be used to construct arbitrarily complex metrics.

\subsection{The Algebra of Metrics}
\label{subsec:metrics}

Formally, a {\bf metric} is a function $f$ that maps the space of 
dataframes $\D$ to a space of arrays, which may be one of:
\begin{itemize}
    \item the space of scalars $\R$ (e.g., sum, mean, count),
    \item the space of vectors $\R^n$ (e.g., distribution), or
    \item the space of matrices $\R^{m \times n}$.
\end{itemize}
We can add, subtract, multiply, and divide metrics $f_1$ and $f_2$ according to the 
general rule 
\[ (f_1 \circ f_2)(D) = f_1(D) \circ f_2(D), \]
assuming that the operation $\circ$ is already well-defined on 
arrays. For example, the churn metric we defined above is really a function of a 
dataframe $D$ that is computed as 
\[ \texttt{churn}(D) = \texttt{Sum("lost")}(D)  /  \texttt{Count("lost")}(D). \]
If the metric returns an array, then operations are typically 
performed element-wise, although operations can be ``broadcast'' 
for arrays of different shapes. \citep{van2011numpy}
If $f_1$ and $f_2$ may produce arrays of incompatible shapes, then 
$(f_1 \circ f_2)(D)$ is NaN.

The space of metrics $\M$ resembles an algebraic field. \citep{dummit2004abstract} 
It is helpful to analogize the space of metrics $\M$ to a field because then we can 
regard adding a new metric as a field extension. Just as extending 
the field of rationals $\Q$ to include $\sqrt{2}$ introduces 
infinitely many new elements of the form 
\[ \{ a + b \sqrt{2}: a, b \in \Q \}, \]
defining a new metric---say, \pyth{SD("x")}---introduces infinitely many 
new metrics of the form 
\[ \{ m_1 + m_2 \cdot \texttt{SD("x")}: m_1, m_2 \in \M \}. \]
In particular, if we take $m_1 = \texttt{Mean("x")}$ and 
$m_2 = \texttt{1.96 / Count("x") ** 0.5}$, then the new metric is familiar as 
the upper bound of a 95\% confidence interval.

\subsection{Operations}

Metrics can also be modified by \pyth{Operation}s. All of the examples 
above involve \pyth{Operation}s:
\begin{enumerate}
    % \item In the baseball example, the \pyth{Count} metric was modified by \pyth{Distribution(over="pitchtype")}. 
    % This operation alters \pyth{Count} in two ways: (1) returns a number for each pitch type instead 
    % of a single number
    % and (2) normalizes these numbers so that they add up to 1.
    % \item In the online retail example, the \pyth{churn} metric (which itself is the composition of 
    % two metrics) was modified by \pyth{PercentChange("experiment", "control")}. This operation 
    % alters \pyth{churn} in two ways: (1) calculates the churn separately for each experiment arm 
    % and (2) reports the percent change in the churn (relative to the control arm). 
    % \item In the minimum wage example, the \pyth{employment} metric was first modified by \pyth{AbsoluteChange("state", "Pennsylvania")},
    % then by \pyth{AbsoluteChange("period", "before"))}. Each has similar effects as the \pyth{PercentChange("experiment", "control")} in last example.
    % Together they alter the \pyth{employment} metric so that the \pyth{Metric} 
    % (1) calculates the average employment rate for every state * period combination;
    % (2) calculates the difference between states (Pennsylvania - New Jersey) for the period before and after the minimum wage increase;
    % (3) calculates the difference of the differences between two periods.
    % \pyth{Operation}s can be chained indefinitely because once taken a \pyth{Metric}, an \pyth{Operation} becomes a function that maps the space of data frames to a space of arrays so is essentially a \pyth{Metric}.
    
    \item In the baseball example, the \pyth{Count} metric is modified by \pyth{Distribution(over="pitchtype")}. This operation has two effects: 
    (1) it returns a value for each pitch type instead of a single aggregate value, 
    and (2) it normalizes these values so that 
    they sum to 1.

    \item In the online retail example, the \pyth{churn} metric (itself a composite of two metrics) is modified by \pyth{PercentChange("experiment", "control")}. 
    This operation similarly alters \pyth{churn} in two ways: (1) it computes the churn rate separately for each experiment arm, and (2) it reports the percent change in churn relative to the control arm.

    \item In the minimum wage example, the \pyth{employment} metric is first modified by \pyth{AbsoluteChange("state", "PA")}, and subsequently by 
    \pyth{AbsoluteChange("period", "before")}.
    Each operation has an effect analogous to \pyth{PercentChange} in the previous example. 
    The net effect is that we:
    (1) compute the average employment rate for each state-period combination, 
    (2) then calculate the difference between states ($\text{PA} - \text{NJ}$) for both the period before and after the minimum wage increase, and
    (3) calculate the difference between periods ($\text{after} - \text{below}$) 
    of those differences.
    The chaining of \pyth{Operation}s is possible because an \pyth{Operation} is itself a \pyth{Metric} that 
    can be modified by another \pyth{Operation}. 
    Therefore, \pyth{Operation}s can be chained indefinitely.
\end{enumerate}

One important class of operations are ``standard error'' operations, which specify a method of 
calculating standard errors. For example, the following code would
produce bootstrap standard errors for the difference-in-differences metric above:
    \begin{python}
(employment_rate
 | AbsoluteChange("STATE_NAME", "PA") 
 | AbsoluteChange("PERIOD", "Before"))
 | Bootstrap()
).compute_on(df)
     \end{python}
Since we often need to quantify uncertainty in metrics, ``standard error'' operations are very 
commonly chained with other operations.

Continuing the analogy to field extensions, we can think of adding an operation 
as analogous to adding the $\sqrt{\cdot}$ operator to $\Q$. This introduces 
not only $\sqrt{2}$, but also $\sqrt{3}$ and $\sqrt{-1}$ to the field.
That is, adding an operation is like a field extension of infinite degree. 
Operations vastly expand the space of possible metrics, far beyond what we could 
hope to achieve by adding new metrics individually.

Operations must modify other metrics; they cannot stand alone as metrics.
Just as the $\sqrt{\cdot}$ operator is not a number, the percent change 
is not a viable metric by itself. We must specify a metric of which to calculate the percent change.

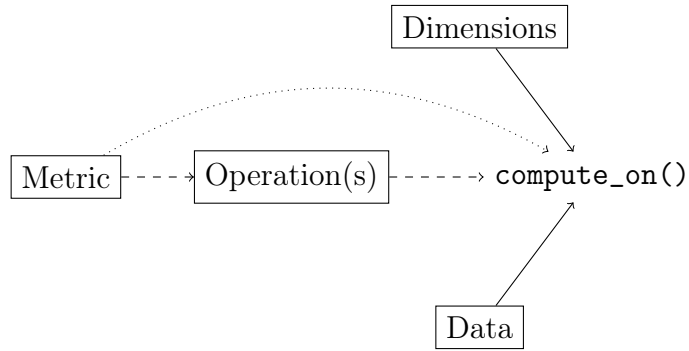
\begin{figure}
    \centering
    \begin{tikzpicture}
    
    \node[rectangle,draw] (metric) at (-4, 0) {Metric};
    \node[rectangle,draw] (op) at (-1, 0) {Operation(s)};
    \node[rectangle,draw] (dim) at (1.5, 2) {Dimensions};
    \node[rectangle,draw] (data) at (1.5, -2) {Data};
    \node (compute) at (3, 0) {\verb\compute_on()\};
    
    \draw[->,dotted,bend left] (metric) edge (compute);
    \draw[->] (dim) -- (compute);
    \draw[->] (data) -- (compute);
    
    \draw[->,dashed] (metric) -- (op);
    \draw[->,dashed] (op) -- (compute);
    
    \end{tikzpicture}
    \caption{The basic flow of a data analysis.}
    \label{fig:data_flow}
\end{figure}

\begin{figure}
\begin{tikzpicture}

\node (input) at (0,0) {
\begin{tabular}{|l|l|l|}
\hline
{\small {\bf region}} & {\small {\bf experiment}}  & {\small {\bf lost}} \\ \hline
US     & control     & 1    \\ \hline
EU     & control     & 0    \\ \hline
US     & treatment1 & 0    \\ \hline
EU     & treatment3 & 1    \\ \hline
US     & control     & 1    \\ \hline
US     & treatment2 & 0    \\ \hline
...    & ...         & ...  \\
\end{tabular}
};

\node[rectangle,draw,align=left] (churn) at (3.9, 3) {\pyth{churn}};
\node[rectangle,draw,align=left] (pct) at (5.3, 0.) {\pyth{churn |} \\ \pyth{PercentChange(...)}};
\node[rectangle,draw,align=left] (bootstrap) at (5.3, -4) {\pyth{churn |} \\ \pyth{PercentChange(...)} \\ \pyth{Bootstrap()}};

\node (churn_output) at (8.5,3) {
\begin{tabular}{|r|}
\hline
churn \\ \hline
0.5    \\  \hline
\end{tabular}
};

\node (pct_output) at (11,0) {
\begin{tabular}{|r|r|}
\hline
experiment   & pct\_change\_of\_churn \\ \hline
control      & 0.0    \\  \hline
treatment1  & 3.2    \\  \hline
treatment2  & 5.3    \\  \hline
treatment3  & 1.8    \\  \hline
\end{tabular}
};

\node (bootstrap_output) at (11.25,-4) {
\begin{tabular}{|r|r|}
\hline
experiment   & se\_pct\_change\_of\_churn \\ \hline
control      & 0.0    \\  \hline
treatment1  & 0.43    \\  \hline
treatment2  & 0.66    \\  \hline
treatment3  & 0.23    \\  \hline
\end{tabular}
};

\draw[->] (input) -- (churn);
\draw[->] (churn) -- (churn_output);
\draw[->] (input) -- (pct);
\draw[->] (pct) -- (pct_output);
\draw[->] (input) -- (bootstrap);
\draw[->] (bootstrap) -- (bootstrap_output);

\end{tikzpicture}
\caption{The input and output data schema for the churn rate computation.}
\label{fig:data_shcema}
\end{figure}
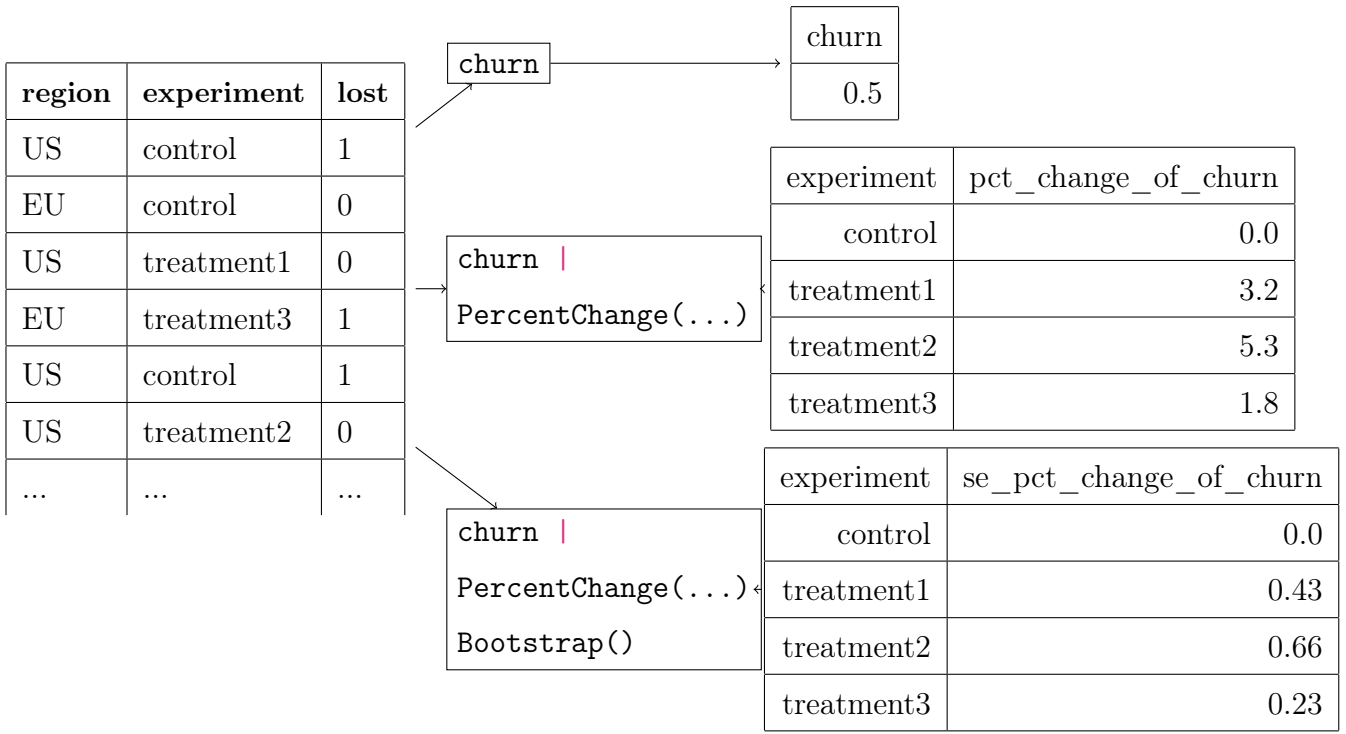

By chaining operations, we can construct arbitrarily complex metrics. 
We can also combine these metrics with different dimensions. To complete the data analysis, 
we finally need to specify a data source. These three components of a data analysis are 
diagrammed in Figure~\ref{fig:data_flow}. The input and output data for the churn rate computation, without any dimension,
are illustrated in Figure~\ref{fig:data_shcema}.

\section{Comparison with Grammars of Data Manipulation}
\label{subsec:comparison}

The idea of calculating metrics over 
dimensions is reminiscent of the 
``split-apply-combine'' framework of \citet{wickham2011split}. 
We first {\em split} the data into many 
smaller data sets, one for each combination of dimensions, called 
a {\bf slice}. Then, we 
{\em apply} the metric to each slice. Finally, we 
{\em combine} the results in the output. 

In many packages \citep{gray1997data,mckinney2010data,wickham2014dplyr}, the 
``split-apply-combine'' framework is implemented using the ``group-by'' operator. 
The user specifies variables to group by and an aggregation function. However, the grouping variables do not 
always correspond neatly to the dimensions, especially 
when metrics are modified by operations. 
To see why, consider the examples from above:

\begin{enumerate}

    \item To implement the baseball example in {\tt dplyr} in {\tt R} (the {\tt pandas} code 
    would be similar), the grouping 
    variables have to include the dimensions {\em and} the pitch types:
    \begin{verbatim}
df %>% group_by(pitcher, batter, pitchtype) %>% 
       count() %>%
       group_by(pitcher, batter) %>%
       mutate(n / sum(n))\end{verbatim}
    The analysis requires two passes through the data. 
    In the first pass, we count at-bats, grouping by the dimensions {\em and} the pitch 
    type. Then, we aggregate again, but this time 
    grouping only by the dimensions, 
    in order to calculate the distribution over the pitch types. 
    
    The problem with this workflow is that it is fragile. If we wanted to 
    add a dimension to this analysis (e.g., \verb\month\), we would need 
    to modify the code in multiple places (i.e., both \verb\group_by\s) to 
    effect this one change. Contrast this with the \Meterstick{} code above, 
    where the dimensions are specified in only one place.

    \item To implement the online retailer example 
    in {\tt dplyr}, we again have to make two passes through the data. 
    \begin{verbatim}
df_by_expt <- df %>% group_by(region, experiment) %>%
                     summarize(churn=sum(lost) / n())
df_treated <- filter(df_by_expt, experiment != "control")
df_control <- filter(df_by_expt, experiment == "control")
df_treated %>% inner_join(df_control,
                          split_by="region",
                          suffix=c("_treated", "_control)) %>%
               mutate(churn_diff=100 * (churn_treated / churn_control - 1))\end{verbatim}

    In the first pass, 
    we group by the dimensions {\em and} the experiment. In the second pass, we 
    join the treated data to the control along the dimensions and calculate the 
    difference. (This generalizes to more than one treatment
    arm.) Again, if we wanted to add a dimension to this 
    analysis, we would need to modify the code in multiple places:
    \verb\group_by()\ and \verb\inner_join()\.
    
    \item To implement the difference-in-differences analysis in {\tt dplyr}, we have to aggregate at the finest level and reshape
    the data multiple times.
    
    \begin{verbatim}
df %>%
  group_by(STATE_NAME, PERIOD) %>%
  summarize(EMP_RATE = mean(EMP, na.rm=T)) %>%
  pivot_wider(names_from = STATE_NAME,
              values_from = EMP_RATE) %>%
  mutate(DIFF=NJ - PA) %>%
  select(PERIOD, DIFF) %>%
  pivot_wider(names_from = PERIOD,
              values_from = DIFF) %>%
  mutate(DIFF_OF_DIFFS=After - Before)
    \end{verbatim}
    
    If we wanted to change the treatment from \verb\STATE_NAME\ to
    something else, we would need to modify the code in three
    places.
    
\end{enumerate}

The code is already quite complex, and we are only calculating 
point estimates. A proper statistical analysis would attach 
standard errors to each of these estimates, which would 
make the code even more complex. 

The problem with a grammar of data manipulation like 
for data analysis is that the syntax does not mirror 
intention; a single component of the analysis corresponds 
to multiple components of code. The above examples 
highlight the ``bottom-up'' nature of data manipulation, 
requiring the analyst to specify low-level details such 
as what variables to group by at each stage. 
By contrast, a grammar of data 
analysis should be ``top-down,'' with its primitives 
matching the conceptual components of the analysis.

On a practical level, the bottom-up code presented 
in this section is ill-suited to the iterative process of 
data science, where questions are constantly being tweaked 
and refined. Data science demands code that is reproducible, 
reusable, and reviewable. If changing one element of the
analysis requires changing multiple lines of code, then it 
becomes difficult to reuse and review the code without 
introducing bugs.

% It is interesting that our primitives correspond to many of the 
% ``Seven Pillars of Statistical Wisdom'' proposed by \citet{stigler2016seven}. 

\section{Implementation}
\label{sec:implementation}

In Section~\ref{sec:grammar}, we described the interface for Meterstick, 
a particular implementation of the grammar of data analysis. In this section, we discuss 
implementation details that further illustrate the recursive and 
composable nature of this grammar.

Before diving into the details, we emphasize that this implementation is separate from the grammar; the analyst does not need to understand these implementation details. Once they specify their intention in the grammar, the library handles the details of carrying out the analysis, whether on a DataFrame or a SQL database. We specify
the implementation details here because we think they 
are of independent interest.

The key idea behind the implementation 
is that every \pyth{Metric}, no matter how complex, 
can be represented as a tree of \pyth{Metric}(s). For example, the 
change in the churn rate with bootstrap standard errors is represented internally 
by the tree shown in Figure~\ref{fig:metric_tree}. An \pyth{Operation} 
is simply a \pyth{Metric} with children.  Each \pyth{Metric}
has methods which calculate the metric on data.

\subsection{DataFrames}

When data is in a DataFrame,
the metric is evaluated using the \pyth{compute_on} method. 
This method specifies how to calculate the metric on a given 
DataFrame over a given set of dimensions. 
The implementation differs depending on whether the 
\pyth{Metric} is an \pyth{Operation} or not.

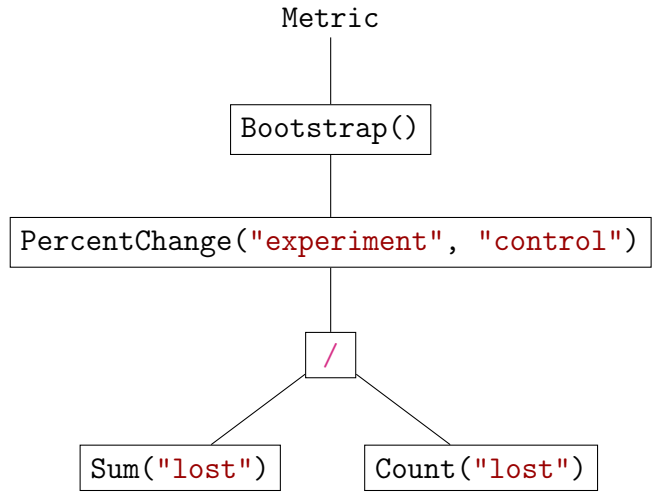
\begin{figure}
    \centering
    \begin{tikzpicture}

    \node (metric) at (0, 0) {\pyth{Metric}};
    
    \node[rectangle,draw] (bootstrap) at (0, -1.5) {\pyth{Bootstrap()}};
    
    \node[rectangle,draw] (pct) at (0, -3) {\pyth{PercentChange("experiment", "control")}};
    
    \node[rectangle,draw] (churn) at (0, -4.5) {\pyth{/}};
    
    \node[rectangle,draw] (sum) at (-2, -6) {\pyth{Sum("lost")}};
    \node[rectangle,draw] (count) at (2, -6) {\pyth{Count("lost")}};
    
    \draw (metric) -- (bootstrap);
    \draw (bootstrap) -- (pct);
    \draw (pct) -- (churn);
    
    \draw (churn) -- (sum);
    \draw (churn) -- (count);
    
    \end{tikzpicture}
    \caption{Example structure of the metric from the online retailer example.}
    \label{fig:metric_tree}
\end{figure}

For simple \pyth{Metric}s which are not \pyth{Operation}s,
such as \pyth{Sum}, the computation can be accomplished by 
simply {\em grouping by} the dimensions. Each type of 
\pyth{Metric} simply needs to implement the \pyth{compute()} method and specify a 
default naming pattern, as shown below.
\begin{python}
class Metric:

  def compute_on(self, data, split_by=None):
    if split_by:
      res = self.compute(data.groupby(split_by))
    else:
      res = [self.compute(data)]
    res = pd.DataFrame(res)
    res.columns = self.names
    return res

  def set_names(self, names):
    self._names = names
    return self

  @property
  def names(self):
    return getattr(self, '_names', self.default_names)

@attrs.define
class Sum(Metric):
  var: str
  
  def compute(self, data):
    return data[self.var].sum()
    
  @property
  def default_names(self):
    return [f'sum_{self.var}']
\end{python}

For \pyth{Operation}s, the computation requires passing data 
{\em down} and results {\em up} the tree. Conceptually, the computation 
can be organized into three steps:
\begin{enumerate}
    \item preprocess the data,
    \item call the \pyth{compute_on} methods of the children on the preprocessed data, and
    \item process the results from the children to compute the final metric.
\end{enumerate}
The general \pyth{Operation} class looks as follows.

\begin{python}
class Operation(Metric):

  def compute_on(self, data, split_by=None):
    data_preprocessed = self.preprocess(data, by)
    child_res = self.compute_children(data_preprocessed, by)
    return self.process_results(child_res, by)
\end{python}
Each \pyth{Operation} simply needs to implement the three methods above. Table~\ref{tab:computation_df} contains 
implementations for the \pyth{Operation}s shown in Figure~\ref{fig:metric_tree}.

\begin{table}[ht]
{\scriptsize
\begin{tabular}{p{.23\textwidth}|p{.25\textwidth}|p{.29\textwidth}|p{.28\textwidth}}
\hline
{\large
\begin{minted}[breaklines]{python}
Operation
\end{minted}
}
% &
% \begin{minted}[breaklines]{python}
% @attrs.define
% class Sum(Metric):
%   var: str
% \end{minted}
&
\begin{minted}[breaklines]{python}
@attrs.define
class Div(Operation):
  child1: Metric
  child2: Metric
\end{minted}
&
\begin{minted}[breaklines]{python}
@attrs.define
class PercentChange(Operation):
  condition: str
  baseline: str
  child: Optional[Metric]=None
\end{minted}
&
\begin{minted}[breaklines]{python}
@attrs.define
class Bootstrap(Operation):
  n_rep: int
  child: Optional[Metric]=None
\end{minted}
\\
\hline
\begin{minted}[breaklines]{python}
def preprocess(
    self, 
    data,
    split_by):
\end{minted}
% &
&
\begin{minted}[breaklines]{python}
return data
\end{minted}
&
\begin{minted}[breaklines]{python}
return data
\end{minted}
&
\begin{minted}[breaklines]{python}
for i in range(self.n_rep):
  yield data.sample(
    frac=1, replace=True)
\end{minted}
\\
% \cline{1-1}
% \cline{3-4}
\begin{minted}[breaklines]{python}
def compute_children(
    self, 
    data, 
    split_by):
\end{minted}
% &
% \begin{minted}[breaklines]{python}
% if not by:
%   return pd.DataFrame(
%       [data[self.var].sum()],
%       columns=[self.name])
% res = data.groupby(split_by)[
%     self.var].sum()
% res.name = self.name
% return pd.DataFrame(res)
% \end{minted}
&
\begin{minted}[breaklines]{python}
return (
  self.child1.compute_on(
      data, split_by),
  self.child2.compute_on(
      data, split_by))
\end{minted}
&
\begin{minted}[breaklines]{python}
return self.child.compute_on(
    data, 
    split_by + [
      self.condition])
\end{minted}
&
\begin{minted}[breaklines]{python}
sample_res = [
  self.child.compute_on(
    sample, split_by) 
  for sample in data]
return pd.concat(
  sample_res, axis=1)
\end{minted}
\\
% \cline{1-1}
% \cline{3-4}
\begin{minted}[breaklines]{python}
def process_results(
    self, 
    child_res, 
    split_by):
\end{minted}
% &
&
\begin{minted}[breaklines]{python}
num, denom = child_res
num.columns = self.names
denom.columns = self.names
return num / denom
\end{minted}
&
\begin{minted}[breaklines]{python}
if split_by:
  base = child_res.xs(
      self.baseline, 
      level=self.condition)
else:
  base = child_res.loc[
    self.baseline]
res = child_res / base - 1
res.columns = self.names
return res * 100
\end{minted}
&
\begin{minted}[breaklines]{python}
std = child_res.T.groupby(
  level=0).std().T
std.columns = self.names
return std
\end{minted}
\\
\begin{minted}[breaklines]{python}
@property
def default_names(self):
\end{minted}
&
\begin{minted}[breaklines]{python}
return map('_div_'.join, 
  zip(self.child1.names, 
      self.child2.names))
\end{minted}
&
\begin{minted}[breaklines]{python}
return [f'pct_change_of_{n}' 
    for n in self.child.names]
\end{minted}
&
\begin{minted}[breaklines]{python}
return [f'se_{n}' 
    for n in self.child.names]
\end{minted}
\\
\hline
\end{tabular}
}
\caption{The computation for each \mintinline{python}{Operation} in Figure~\ref{fig:metric_tree} requires three steps.}
\label{tab:computation_df}
\end{table}

\subsection{SQL}

It is not always possible to analyze data in DataFrames. For example, 
the size of the data set may exceed computer memory. 
In these situations, data is typically stored in relational databases and analyzed using SQL queries. Therefore, 
each \pyth{Metric} also has a 
\pyth{to_sql} method that 
generates a SQL query which calculates the metric.\footnote{The SQL dialect in the following implementation is GoogleSQL.} 
% \pyth{compute_on_sql} method whose signature is similar to that of, mirroring the familiar structure of its \pyth{compute_on} counterpart.

% The implementation of \pyth{compute_on_sql} vary depending on the specific \pyth{Metric}. If a \pyth{Metric} can be entirely translated into SQL, it will incorporate a \pyth{to_sql} method. 
% This method crafts the precise SQL query needed for the \pyth{Metric}, which is then executed within the \pyth{compute_on_sql} function.

As before, the implementation differs according
to whether the \pyth{Metric} is an \pyth{Operation} or not.
For simple \pyth{Metric}s, the computation can be carried out in a single 
query. Each type of \pyth{Metric} needs only to specify a
SQL expression that computes it, as shown below.

\begin{python}
class Metric:
  
  def sql_aggregate(self, data, dimensions):
    # Helper function for constructing aggregation queries.
    dim_sql = ','.join(dimensions) + ',' if dimensions else ''
    groupby = f'GROUP BY ' + ','.join(dimensions) if dim_sql else ''
    val_cols = ','.join([f'{s} AS {n}' 
                         for s, n in zip(self.sql, self.names)])
    return f'SELECT {dim_sql} {val_cols} FROM {data} {groupby}'

  def to_sql(self, data, split_by=None):
    return self.sql_aggregate(data, split_by)

@attrs.define
class Sum(Metric):
  var: str
  
  @property
  def sql(self):
    return [f'SUM({self.var})']
\end{python}

For \pyth{Operation}s, we organize SQL 
generation into the same
three steps (preprocess, compute on children, and 
process results) as for DataFrames, except that each 
step now returns a SQL subquery instead of a DataFrame. 
Because each child may add new dimensions 
to the data, these dimensions are tracked in the 
property \pyth{extra_dims}. 
(This was not necessary for DataFrames because these dimensions could 
be stored directly in the index of the DataFrame itself.) 

\begin{python}
class Operation(Metric):
  
  def sql_select(self, data, dimensions):
    # Helper function for constructing select queries.
    dim_sql = ','.join(dimensions) + ',' if dimensions else ''
    val_cols = ','.join([f'{s} AS {n}'
                         for s, n in zip(self.sql, self.names)])
    return f'SELECT {dim_sql} {val_cols} FROM {data}'

  def to_sql(self, data, split_by=None):
    data_preprocessed = self.preprocess_sql(data, split_by)
    children_query = self.children_to_sql(data_preprocessed, split_by)
    return self.assemble_query(children_query, split_by)

  @property
  def extra_dims(self):
    return []
\end{python}
Each \pyth{Operation} simply needs to implement the three methods in \pyth{to_sql}, as well as
the \pyth{extra_dims} property. Table~\ref{tab:to_sql} contains 
implementations for the \pyth{Operation}s in Figure~\ref{fig:metric_tree}.

\begin{table}[ht]
{\scriptsize
\begin{tabular}{p{.21\textwidth}|p{.24\textwidth}|p{.304\textwidth}|p{.28\textwidth}}
\hline
{\large
\begin{minted}[breaklines]{python}
Operation
\end{minted}
}
&
\begin{minted}[breaklines]{python}
@attrs.define
class Div(Operation):
  child1: Metric
  child2: Metric
\end{minted}
&
\begin{minted}[breaklines]{python}
@attrs.define
class PercantChange(Operation):
  condition: str
  baseline: str
  child: Optional[Metric]=None
\end{minted}
&
\begin{minted}[breaklines]{python}
@attrs.define
class Bootstrap(Operation):
  n_rep: int
  child: Optional[Metric]=None
\end{minted}
\\
\hline
\begin{minted}[breaklines]{python}
def preprocess_to_sql(
    self, data, 
    split_by):
\end{minted}
&
\begin{minted}[breaklines]{python}
return data
\end{minted}
% &
% \begin{minted}[breaklines]{python}
% return data
% \end{minted}
&
\begin{minted}[breaklines]{python}
return data
\end{minted}
&
\begin{minted}[breaklines]{python}
# resample_n_times is defined
# in the Appendix due to space
return resample_n_times(
  data, split_by, self.n_rep)
\end{minted}
\\
% \cline{1-1}
% \cline{3-4}
\begin{minted}[breaklines]{python}
def children_to_sql(
    self, data,
    split_by):
\end{minted}
&
\begin{minted}[breaklines]{python}
return data
\end{minted}
&
\begin{minted}[breaklines]{python}
return self.child.to_sql(
  data, split_by + [
    self.condition])
\end{minted}
&
\begin{minted}[breaklines]{python}
return (*data, 
  self.child.to_sql('Samples', 
   split_by + ['sample_idx']))
\end{minted}
\\
% \cline{1-1}
% \cline{3-4}
\begin{minted}[breaklines]{python}
def assemble_query(
    self, 
    child_res, 
    split_by):
\end{minted}
&
% &
\begin{minted}[breaklines]{python}
return self.sql_aggregate(
  child_res, split_by)
\end{minted}
&
\begin{minted}[breaklines]{python}
dims = self.extra_dims + split_by
u = ','.join(dims[1:])
join = f'T JOIN Base USING ({u})'
if not u:
  join = 'T CROSS JOIN Base'
return f'''
WITH T AS ({child_res}),
Base AS (SELECT *
EXCEPT ({self.condition}) FROM T
WHERE {self.condition}
   = '{self.baseline}')
{self.sql_select(join, dims)}'''
\end{minted}
&
\begin{minted}[breaklines]{python}
(input_data, samples, 
  sample_res) = child_res
sql = self.sql_aggregate(
  'SampleRes', 
  by + self.extra_dims)
return f'''
  CREATE TEMP TABLE Data 
    AS ({input_data});
  WITH Samples AS ({samples}), 
  SampleRes AS ({sample_res})
  {sql}'''
\end{minted}
\\
\hline
\begin{minted}[breaklines]{python}
@property
def sql(self):
\end{minted}
% &
% \begin{minted}[breaklines]{python}
% return f'SUM({self.var})'
% \end{minted}
&
\begin{minted}[breaklines]{python}
return map(' / '.join, 
  zip(self.child1.sql, 
      self.child2.sql))
\end{minted}
&
\begin{minted}[breaklines]{python}
return [
  f'(T.{c} / Base.{c} - 1) * 100'
  for c in self.child.names]
\end{minted}
&
\begin{minted}[breaklines]{python}
return [f'STDDEV({n})' 
  for n in self.child.names]
\end{minted}
\\
\hline
\begin{minted}[breaklines]{python}
@property
def extra_dims(self):
\end{minted}
&
\begin{minted}[breaklines]{python}
return []
\end{minted}
&
\begin{minted}[breaklines]{python}
return ([self.condition] + 
         self.child.extra_dims)
\end{minted}
&
\begin{minted}[breaklines]{python}
return self.child.extra_dims
\end{minted}
\\
\hline
\end{tabular}
}
\caption{The SQL generator for the \mintinline{python}{Operation}s in Figure~\ref{fig:metric_tree}.}
\label{tab:to_sql}
\end{table}

\subsection{Summary of Features}

We have already seen how a grammar of data analysis 
provides flexibility and adaptability as needs evolve, 
and simplifies code review and maintenance. The 
Meterstick implementation of this grammar offers 
many built-in \pyth{Metric}s, such as
\pyth{Sum}, \pyth{Mean}, \pyth{Quantile}, 
and \pyth{Variance}, and \pyth{Operation}s, such as
\pyth{Distribution}, \pyth{AbsoluteChange}, 
\pyth{Jackknife} and \pyth{Bootstrap}.
It allows these operations to be chained 
indefinitely, as specified by the grammar. 
Moreover, Meterstick allows users to define their 
own metrics (by implementing the methods described 
above) and chain them with the built-in \pyth{Operation}s 
to produce even more analysis pipelines. One common 
use case is that a data scientist develops a new 
metric and wants to calculate the metric, 
along with a confidence interval. If they implement 
the metric in Meterstick, then they get a
bootstrap or jackknife confidence interval for free.

Meterstick is also able to abstract away the 
implementation details, especially efficiency optimizations, 
from the analyst. For example, 
when calculating multiple metrics,
intermediate metrics may be shared across multiple 
metrics. In this situation, Meterstick caches the 
values of these intermediate 
metrics so that they can be reused, speeding up
the computation.

Finally, Meterstick implements the grammar in a 
backend-agnostic way, meaning that the same \pyth{Metric}
can calculated on a DataFrame or a SQL database by 
calling \pyth{compute_on} or \pyth{to_sql}, 
respectively. This allows analyses to be prototyped
on smaller data sets and scaled seamlessly to data warehouses.

\section{Conclusion}
\label{sec:conc}

A grammar of data analysis must sit at a higher level of abstraction than a grammar
of data manipulation, in order to represent common patterns of data analysis. In 
this paper, we have proposed a grammar that can flexibly describe any analysis that can be characterized as 
``calculating metrics over dimensions,'' in a way that allows frequent iteration over different metrics and different dimensions. We have seen how existing grammars of data manipulation fail 
to provide the necessary flexibility and modularity for these types of analyses. We have demonstrated how to 
implement this grammar in a backend-agnostic way so that
it can be applied just as easily to small DataFrames in
memory as to databases in the cloud.

This grammar of data analysis is built upon the idea that metrics can be composed and 
modified to produce more complex metrics. By providing a common language for expressing intention, it can improve communication and collaboration among analysts, promote best practices, and facilitate data exploration and metric development. Moreover, the grammar's focus on modularity and reusability can lead to more efficient and maintainable analysis pipelines.
\appendix

\begin{center}
{\large\bf SUPPLEMENTARY MATERIAL}
\end{center}

\begin{description}

\item[Meterstick package] An open-source implementation of this grammar of data analysis, 
as described in \ref{sec:implementation}, is available at \url{https://www.github.com/google/meterstick}.

\end{description}

\section{Bootstrap Implementation}
\label{sec:appendix}
Below is the implementation of \pyth{resample_n_times} that is used in \pyth{Bootstrap}.
\begin{python}
def resample_n_times(data, split_by, n_rep):
  by_sql = ','.join(split_by) + ',' if split_by else ''
  input_data = f'''
    SELECT
      *,
      ROW_NUMBER() OVER (PARTITION BY sample_idx) AS row_number,
      CEILING(RAND() * COUNT(*) OVER (PARTITION BY sample_idx))
        AS random_row_number,
    FROM {data},
    UNNEST(GENERATE_ARRAY(1, {n_rep})) AS sample_idx'''
  samples = f'''
    SELECT b.*
    FROM (
      SELECT
        {by_sql}
        sample_idx,
        random_row_number AS row_number
      FROM Data) AS a
    JOIN Data AS b
    USING ({by_sql} sample_idx, row_number)'''
  return (input_data, samples)
\end{python}

\bibliographystyle{agsm}

\bibliography{Bibliography-MM-MC}
\end{document}